\documentclass[aps,prd,showpacs,nobibnotes,twocolumn,preprintnumbers,
nobalancelastpage,superscriptaddress]{revtex4}
\usepackage{graphicx}
\usepackage{latexsym}
\usepackage{dcolumn}
\usepackage{bm}
\input{epsf}
\newcommand{\be}{\begin{equation}}
\newcommand{\ee}{\end{equation}}
\newcommand{\ba}{\begin{eqnarray}}
\newcommand{\ea}{\end{eqnarray}}
\newcommand{\no}{\nonumber}
\newcommand{\bfi}{\begin{figure}
\epsfxsize=8cm
\epsffile}
\newcommand{\efi}{\end{figure}}
\newcommand{\bi}{\begin{itemize}}
\newcommand{\ei}{\end{itemize}}
\newcommand{\la}{\lesssim}

\newcommand{\mpch}{h^{-1} {\rm Mpc}}
\newcommand{\gpch}{h^{-1} {\rm Gpc}}

\newcommand{\dif}{\mathrm{d}}

\begin{document}

\title{Gaussianizing the non-Gaussian lensing convergence field I: the performance of
the Gaussianization}
\author{Yu Yu}
\email{yuyu22@shao.ac.cn}
\affiliation{Key laboratory for research in galaxies and cosmology, Shanghai Astronomical Observatory, Chinese Academy of
 Science, 80 Nandan Road, Shanghai, China, 200030}
\author{Pengjie Zhang, Weipeng Lin}
\affiliation{Key laboratory for research in galaxies and cosmology, Shanghai Astronomical Observatory, Chinese Academy of
  Science, 80 Nandan Road, Shanghai, China, 200030}
\author{Weiguang Cui}
\affiliation{Key laboratory for research in galaxies and cosmology, Shanghai Astronomical Observatory, Chinese Academy of
  Science, 80 Nandan Road, Shanghai, China, 200030}
\affiliation{Astronomy Unit, Department of Physics, University of Trieste, Tiepolo 11, I-34131 Trieste, Italy}
\author{James N. Fry}
\affiliation{Department of Physics, University of Florida, Gainesville Florida 32611-8440, USA}
\begin{abstract}
Motivated by recent works of Neyrinck et al. 2009 and Scherrer et al. 2010,
we proposed a Gaussian transformation to Gaussianize the non-Gaussian lensing convergence field $\kappa$.
It performs a local monotonic transformation $\kappa\rightarrow y$ pixel by pixel
to make the fine-scale one-point probability distribution function of the new variable $y$ Gaussian.
We tested whether the whole $y$ field is Gaussian through N-body simulations.
(1) We found that the proposed Gaussianization suppresses the non-Gaussianity by orders of magnitude,
in measures of the skewness, the kurtosis, the 5th- and 6th-order cumulants of the $y$ field
smoothed over various angular scales, relative to that of the corresponding smoothed $\kappa$ field.
The residual non-Gaussianities are often consistent with zero within the statistical errors.
(2) The Gaussianization significantly suppresses the bispectrum.
Furthermore, the residual scatters about zero, depending on the configuration in the Fourier space.
(3) The Gaussianization works with even better performance for the 2D fields of the matter density
projected over $\sim 300 \mpch$ distance interval centered at $z\in(0,2)$,
which can be reconstructed from the weak lensing tomography.
(4) We identified imperfectness and complexities of the proposed  Gaussianization.
We noticed weak residual non-Gaussianity in the $y$ field.
We verified the widely used logarithmic transformation as a good approximation to the Gaussian transformation.
However, we also found noticeable deviations.

\end{abstract}
\maketitle
\section{Introduction}
\label{sec:introduction}

Weak gravitational lensing has been established as one of the most powerful
probes of the dark universe
\citep{Bartelmann01,Refregier03,Albrecht06,Munshi08,Hoekstra08}.
Meanwhile, it is  facing many challenges. One of them is the nonlinear
evolution of the large scale structure (LSS). An inevitable consequence is the
significant presence of non-Gaussianity in the weak lensing
field at subdegree scales, which has been confirmed by observations
\cite{skewnessmeasurements}. For this reason, the most widely used statistics,
the power spectrum, fails to capture all the cosmological information and a significant
amount of information is encoded in the three point lensing statistics
\cite{NGtheory} and  perhaps also in even higher-order statistics. It also
increases cosmic variance in the power spectrum, introduces correlated errors
between different modes and complicates the error analysis based on the power
spectrum measurement \cite{Rimes05,Rimes06,Semboloni07,Lee08,Takada09}.

A number of recent works on the 3D matter density field lead us
to propose a possible solution to the lensing non-Gaussianity problem.
Neyrinck et al. 2009 \cite{Neyrinck09} showed that a local logarithmic
transformation $\delta\rightarrow \ln(1+\delta)$
is able to significantly improve the cosmological information encoded in the 3D power spectrum of the new field.
Since the transformation is nonlinear, it compresses information encoded in higher-order statistics
of the original density field into the power spectrum of the new field.
This significant improvement in the cosmological information also implies that
the new field is significantly more Gaussian.

Scherrer et al. 2010  \cite{Scherrer10} introduced a novel and powerful mathematical tool,
the copula, to study the statistics of the LSS.
The n-point copula is the function that relates the one-point CDF (cumulative distribution function) to the joint n-point CDF.
A mathematical theorem states that the copula always exist and must be unique (refer to \cite{Scherrer10} and references therein for details).
Hence the one-point PDF (probability distribution function) and the full n-point copula ($n=2,3\cdots$) present a complete description of a field.
A surprising finding by \cite{Scherrer10} through N-body simulation is that,
despite strong  non-Gaussianity in the 3D matter distribution, the two-point copula is
virtually Gaussian.  This motivates the authors to postulate the Gaussian
copula hypothesis, that all n-point copula functions are Gaussian.

Copula is invariant under monotonic transformation.
Hence the finding above implies that, if we find a local monotonic transformation to make the one-point PDF Gaussian,
other statistics (n-point PDFs and equivalently n-point correlation functions) will be Gaussian.
The whole field will then be Gaussian and hence cosmological information otherwise encoded in higher-order statistics
is now compressed into the power spectrum.
This helps to understand the enhancement of cosmological information in the power spectrum
by performing a logarithmic transformation (\cite{Neyrinck09} and also in their more recent work \cite{Neyrinck10}).
Since the one-point PDF of the 3D density distribution is know to be close to log-normal,
the logarithmic transformation makes the one-point PDF nearly Gaussian.
Along with the nearly Gaussian copula as found by \cite{Scherrer10}, the logarithmic transformation should indeed work well
to Gaussianize the 3D density field and compress cosmological information otherwise encoded in higher-order statistics into the power spectrum \cite{footnote1}.

These studies focus on the 3D matter distribution.
Nevertheless, this naturally motivates us to propose the following postulation on the 2D lensing field.
Without loss of generality, the lensing field is described by the lensing convergence $\kappa$ throughout the paper.
Our postulation, which is a natural and straightforward extension of the Gaussian copula hypothesis proposed in \cite{Scherrer10},
then states that,
\bi
\item {\it Local monotonic Gaussian transformation $\kappa\rightarrow y$
  makes the whole $y$ field (effectively) Gaussian.}
 \ei
Here, local monotonic Gaussian transformations are the transformations $\kappa\rightarrow y$
that make the one-point PDF of the pixelized field Gaussian with rank order preserved.
The grid size for pixelization should be suitable to extract cosmological information.
The scale should be enough to resolve nonlinear regime, but meanwhile the shot noise should not dominate the statistics.
We call the field binned into a uniform grid suitable to extract cosmological information as {\it fine-scale} field.
A field being (effectively) Gaussian means that all the measures of non-Gaussianity are (significantly suppressed and) consistent with zero within the error bars,
under the consideration on the existence of fluctuations in statistics due to a finite volume.
If this postulation is indeed correct, all cosmological information is encoded in the transformation $\kappa\rightarrow y$ and the power spectrum of $y$.
This would significantly simplify extracting cosmological parameters from the lensing field.

We tested this postulation using our numerical simulations.
The measures of non-Gaussianity that we investigate include the cumulants up to 6th-order of the {\it smoothed} $y$ field
and the bispectrum of the {\it fine-scale} $y$ field.
By definition, the fine-scale $y$ field derived by pixel-to-pixel mapping has Gaussian cumulants due to the forced Gaussianization.
However, this does not apply to the smoothed one, since by smoothing over several pixel scale, pixel-pixel correlation comes into the cumulants.
We found that, for all these measures, the non-Gaussianity is suppressed by orders of magnitude and the residual non-Gaussianity is often negligible.
From a practical viewpoint, these results, though limited to statistics no higher than 6th-order, prove the above postulation.
Hence it shows that the non-Gaussian lensing field can be indeed Gaussianized by a simple local variable transformation.
Although all these tests are carried out against the 2D convergence field,
it strongly supports the original Gaussian copula hypothesis on the 3D matter distribution \cite{Scherrer10}.

Our Gaussianization proposal is closely related to, but not the same as,
the previous  Gaussianization  proposed to recover the primordial
density distribution (e.g. \cite{Weinberg92,fengll00}). Such proposal
assumes an one-to-one monotonic correspondence between the observed density
(matter,  galaxy, Lyman-$\alpha$ flux, etc.) distribution and the primordial
one.  Our Gaussianization  relies on the empirical finding of
\cite{Scherrer10}. However, it does not rely on the above assumption and can
work surprisingly well even in the deeply nonlinear regime where this
assumption breaks. But for the same reason, the Gaussianized field is not
necessarily closely related to the primordial density distribution.

Our work complements several recent works, but through different and independent approaches.
(1) \cite{Seo10} found that a logarithmic transformation $\kappa\rightarrow \ln(1+\kappa/\kappa_0)$ can significantly increase the information
contained in the lensing power spectrum while reducing correlations between different multipole $\ell$ modes.
This work focused on the two-point statistics (and its variance).
Our work focuses on the direct measures of the non-Gaussianity, such as the skewness and the bispectrum.
Both works are hence highly complementary and consistent, both supporting the feasibility of Gaussianizing the non-Gaussian lensing field.
(2) \cite{TJZhang10,HRYu10} took a different approach to Gaussianize the 3D matter density field and the 2D lensing convergence field.
By using nonlinear wavelet Weiner filtering, they are able to remove/suppress small scale non-Gaussian structure,
rendering the fields more Gaussian and hence increasing the information contained in the 2D and 3D power spectra.
(3) \cite{boxcox} also investigated the information on cosmology contained in Gaussianized weak gravitational lensing convergence fields,
when our work is under modification to publish.
By employing Box-Cox transformations to determine optimal transformations to Gaussianity,
they developed analytical models for the transformed power spectrum, including effects of noise and smoothing.
They found that optimized Box-Cox transformations perform better than logarithmic transformation,
but both yield very similar results for the signal-to-noise and parameter constraints in ideal case.
When adding a realistic level of shape noise, all transformations perform poorly.

Our paper is organized as follows: In \S \ref{sec:Gaussianization} we briefly introduce the weak lensing basics and propose the Gaussianization method.
Measures of the non-Gaussianity and hence measures of the Gaussianization performance are also introduced in this section.
The simulation used to construct weak lensing convergence maps is described in \S \ref{sec:lensingmap}.
The measures of the non-Gaussianity in simulated lensing convergence maps are presented in \S \ref{sec:kappa}.
In \S \ref{sec:sigma} we repeat the analysis for the 2D matter density field projected over the box size,
which corresponds to the redshift resolved matter field reconstructed from the weak lensing tomography.
In \S \ref{sec:conclusion} we outline key issues for further investigation.


\section{Gaussianizing the lensing convergence field}
\label{sec:Gaussianization} The weak lensing basics can be found in review articles
\citep{Bartelmann01,Refregier03,Albrecht06,Munshi08,Hoekstra08} and a textbook \cite{ModernCosmology}.
Under the Born approximation, the weak lensing convergence $\kappa$ of a source at redshift $z_s$ and
direction $\hat{n}$  can be expressed as \cite{Refregier03}
\be
\label{eqn:kappa} \kappa(\hat{n},z_s)=\int
\delta_m(\hat{n},z)W(z,z_s)d\tilde{\chi}\ .
\ee
Here, $\delta_m$ is the matter overdensity.
$\chi\equiv \chi(z)$ is the comoving angular diameter distance to the lens redshift $z$.
We conveniently express $\chi$ in units of the Hubble radius, $\tilde{\chi}\equiv \chi/(c/H_0)$,
in which $H_0$ is the Hubble constant today.
The lensing kernel $W(z,z_s)$ for a source at redshift $z_s$ and a lens at redshift $z$ is given by
\be
W(z,z_s)=\frac{3}{2}\Omega_m (1+z) \tilde{\chi}(z)\left[1-\frac{\chi(z)}{\chi(z_s)}\right]\ .
\ee
when $z\leq z_s$ and zero otherwise.
Here $\Omega_m$ is the cosmological matter density of the universe in units of the critical density.
The above expression is valid for the flat cosmology we consider throughout the paper.

Weak lensing directly probes the matter overdensity projected along the line-of-sight from the observer to the source galaxies.
Through the $z_s$ dependence in $\kappa$, the lensing tomography works to reconstruct the 3D matter  distribution, namely $\delta_m(\hat{n},z)$.
In reality, due to limited source redshifts, measurement errors and the relatively wide and smooth lensing kernel,
the reconstruction is most feasible for the binned matter overdensity $\delta^{\Sigma}$, namely, the matter column density averaged over a narrow redshift bin.
The one for the $i$-th redshift bin ($z_i-\Delta z_i/2<z<z_i+\Delta z_i/2$) is defined as
\be
\delta^\Sigma_i(\hat{n})\equiv \frac{\int_{z_i-\Delta z_i/2}^{z_i+\Delta
    z_i/2}\delta_m(\hat{n},z)d\chi}{\Delta \chi_i}\ .
\ee
Here $\Delta \chi_i$ is  the comoving width of the $i$-th redshift bin,
\be
 \Delta \chi_i\equiv \chi\left(z_i+\frac{\Delta
   z_i}{2}\right)-\chi\left(z_i-\frac{\Delta z_i}{2}\right)\ .
\ee
Eq. \ref{eqn:kappa} can then be rewritten as
\be
\label{eqn:kappasum}
\kappa(\hat{n})\simeq \sum_i \delta^\Sigma_i(\hat{n}) W_i\Delta \chi_i\ ,
\ee
where the weight $W_i\equiv W(z_i,z_s)$.
Eq. \ref{eqn:kappasum} is exact in the limit $\Delta \chi_i\rightarrow 0$ such that we can neglect the variation of $W$ over the given redshift bin.
In reality, since the typical width of the lensing kernel $W$ is comparable to the Hubble radius,
Eq. \ref{eqn:kappasum} is an excellent approximation as long as $\Delta
\chi_i\ll c/H_0=3\gpch$.

Both the $\kappa(\hat{n})$ and $\delta_i^{\Sigma}(\hat{n})$ fields are, in principle, observable.
$\delta_i^{\Sigma}$ is a redshift resolved version of $\kappa$.
It contains more detailed information on the LSS evolution and hence the dark universe.
For subdegree angular scales of interest and $O(100\mpch)$ bin width,  $\delta^\Sigma_i$ each are effectively uncorrelated with each other.
Understanding the statistics of $\delta^\Sigma_i$ then allows us
to predict the statistics of $\kappa$ through Eq. \ref{eqn:kappasum} in a straightforward fashion.

Both $\kappa(\hat{n})$ and $\delta^{\Sigma}(\hat{n})$ are nonlinear
and non-Gaussian, meaning that the two-point angular correlation
function (and equivalently the power spectrum) fails to capture all
cosmological information encoded in these fields.  A straightforward
remedy for such information loss is to go up through the n-point
correlation hierarchy. However calculating three-point and higher-order
statistics are daunting for large data sets. Even worse, it is
not clear where the information saturates. The Gaussianization on
$\kappa$ and $\delta_i^{\Sigma}$ provides an alternative. If
succeed,  it  will truncate the analysis at the level of the two-point
statistics and hence significantly simplify the analysis.

\subsection{The Gaussianization procedure}
\label{subsec:procedure}
We consider an idealized case that the lensing data is binned into a uniform grid suitable to extract cosmological information.
For each pixel centered at direction $\hat{n}$, we measure $\kappa(\hat{n})$ (and $\delta_i^{\Sigma}(\vec{r}_{\perp})$,
where $\vec{r}_\perp$ is the coordinate perpendicular to the line-of-sight).
We call it a {\it fine-scale} lensing convergence (and $\delta^\Sigma$) map.
From such data we can directly measure the one-point PDF of $\kappa$ and $\delta^\Sigma$.
These PDFs are non-Gaussian, for subdegree pixel size.

One can always find a set of monotonic local (pixel by pixel) transformations $\kappa\rightarrow y$ ($\delta^\Sigma\rightarrow y$)
such that the one-point PDF of the {\it fine-scale} $y$ field is Gaussian.
The numerical solution can be found by the following equations,
\be
\label{eqn:kappay}
\int_{-\infty}^{y}P_G(y){\dif}y=\int_{-\infty}^{\kappa}P(\kappa){\dif}\kappa\ ,
\ee
and
\be
\label{eqn:deltay}
\int_{-\infty}^{y}P_G(y){\dif}y=\int_{-\infty}^{\delta^\Sigma}P(\delta^\Sigma){\dif}\delta^\Sigma\ .
\ee
Here by definition $P_G(y)=\exp(-y^2/2\sigma_y^2)/\sqrt{2\pi}\sigma_y$ is the Gaussian PDF and $\sigma_y$ is the  rms dispersion of $y$.

Throughout the paper, we call the solution to the above equations as {\it the Gaussian transformation}.
It is the transformation that we propose to Gaussianize the whole lensing convergence field.
Such transformations are unique, up to a scaling $y\rightarrow a y$, where $a\propto \sigma_y$ reflects the freedom in choosing $\sigma_y$.
Such scaling does not affect the non-Gaussianity of the $y$ maps.
By choosing a suitable scaling, the $y$ field can have other desirable properties. We will briefly discuss this issue later in \S \ref{sec:conclusion}.

\subsection{Measures of the Gaussianization performance}
The performance of the proposed Gaussianization procedure can be quantified by various non-Gaussianity measures.
We have investigated the cumulants up to the 6th order and the reduced bispectrum.

(1) The $n$th-order cumulants of the {\it smoothed} $y$ field, denoted as $K_n(y_S)$.
These are nontrivial checks, despite the fact that $n$th-order cumulants of the fine-scale $y$ field are Gaussian by definition.
Smoothing introduces pixel-pixel correlation and can give rise to non-Gaussian $y_S$ cumulants.
We adopt the conventional definition of the normalized cumulants in statistics.
\ba
\label{eqn:cumulants}
K_3&\equiv& \frac{\langle y_S^3\rangle}{\langle y_S^2\rangle^{3/2}}\no \ ,\\
K_4&\equiv& \frac{\langle y_S^4\rangle}{\langle y_S^2\rangle^{2}}-3\no \ ,\\
K_5&\equiv& \frac{\langle y_S^5\rangle}{\langle
  y_S^2\rangle^{5/2}}-10\frac{\langle y_S^3\rangle}{\langle
  y_S^2\rangle^{3/2}}\ ,\\
K_6&\equiv& \frac{\langle y_S^6\rangle}{\langle
  y_S^2\rangle^{3}}-10\frac{\langle y_S^3\rangle^2}{\langle
  y_S^2\rangle^{3}}-15 \frac{\langle y_S^4\rangle}{\langle
  y_S^2\rangle^{2}}+30 \no\  .
\ea
The cumulants before the Gaussianization is defined by replacing $y_S$ with $\kappa_S$ (and $\delta^\Sigma_S$) smoothed in the same way.

We caution that $K_n$ differ from the cumulants $S_n$ usually adopted in LSS study (e.g. in the review article \cite{Bernardeau02}).
For example, $S_3\equiv \langle y_S^3\rangle/\langle y_S^2\rangle^2$.
There are two major reasons to define the cumulants $K_n$ according to Eq. \ref{eqn:cumulants}.
(i) $S_n$ are not invariant under the transformation $y\rightarrow a y$, while the non-Gaussianity is obviously invariant under this transformation.
Since we will quantify the Gaussianization performance by comparing the cumulants prior to and posterior to the Gaussianization,
the usual definitions $S_n$ could mislead us.
On the contrary, $K_n$ defined above are invariant under the transformation $y\rightarrow a y$ and hence allows for unbiased comparison.
(ii) $K_n$ itself is a straightforward measure of the Gaussianization performance, since $|K_n|\ll 1$ means weak non-Gaussianity.
On the contrary, $S_n$ does not have this property, again due to its dependence on the scaling $y\rightarrow ay$.

(2) The redefined reduced bispectrum $q$.
For the 2D $\delta^\Sigma$ fields, we define
\be
\label{eqn:Q}
q(\vec{k}_{\perp,1},\vec{k}_{\perp,2},\vec{k}_{\perp,3})=\frac{B(\vec{k}_{\perp,1},\vec{k}_{\perp,2},\vec{k}_{\perp,3})}{\left[P(k_{\perp,1})P(k_{\perp,2})+{\rm
      cyc.}\right]^{3/4}}\ .
\ee
Here, $\vec{k}_{\perp,i}$ ($i=1,2,3$) are 2D wavevectors and $\sum_i \vec{k}_{\perp,i}=0$.
The bispectrum $B(\vec{k}_{\perp,1},\vec{k}_{\perp,2},\vec{k}_{\perp,3})=
\langle \delta^\Sigma(\vec{k}_{\perp,1})\delta^\Sigma(\vec{k}_{\perp,2})\delta^\Sigma(\vec{k}_{\perp,3})\rangle$
and the power spectrum $P(k_{\perp})=\langle |\delta^\Sigma(\vec{k}_{\perp})|^2\rangle$.
For the lensing convergence  maps, we replaced $\vec{k}_{\perp,i}$ with the 2D multipole vector $\vec{\ell}$.

The $q$ that we define has a power index $3/4$ in the denominator, which differs from the power index $1$ in the capital $Q$ widely used in  the LSS literature.
Although the power index $1$ is the natural choice preferred by the 2nd-order  perturbation theory, it causes $Q$ to vary under the transformation $y\rightarrow ay$.
For this reason, $Q$ is not suitable to describe the Gaussianization performance.
Conversely, $q$ is invariant under the transformation $y\rightarrow ay$.
Unless otherwise specified, we refer to $q$ as the reduced bispectrum throughout the paper.

Here we remind readers that the Gaussian transformation does not reduce higher-order statistics in all situations.
Gaussianization has no effect on a field only containing 0's and 1's,
which arises observationally if too small a cell size is used for galaxy number counting field.
Only the field with all n{\it th} order copula being Gaussian can be fully Gaussianized with this Gaussianization procedure.


\section{Simulated lensing maps}
\label{sec:lensingmap}
We numerically evaluate $K_n$ and $q$ and test the Gaussianization performance using simulated lensing maps.
Our N-body simulations were run using the Gadget-2 code \cite{Gadget2}.
They all have box size $L=300\mpch$ and adopt the standard $\Lambda$CDM cosmology,
with $\Omega_m=0.266$, $\Omega_{\Lambda}=1-\Omega_m$, $\sigma_8=0.801$, $h=0.71$ and $n_s=0.963$.
More details can be found in \cite{Cui10}.
The maps are generated as follows.
(1) For each output, we project along $x$, $y$ and $z$ directions respectively to generate three maps of
$\delta^\Sigma_i(\vec{r}_{\perp})$, where $\vec{r}_{\perp}$ is the 2D coordinate in comoving distance.
(2) We then convert $\vec{r}_{\perp}$ to the angular coordinate through $\hat{n}=\vec{r}_{\perp}/\chi(z_i)$,
where $z_i$ is the redshift of the corresponding output.
(3) We have chosen the output redshifts such that any two adjacent outputs are separated by the box size $L=300\mpch$ in comoving distance.
So by stacking maps of $\delta^{\Sigma}_i$ according to Eq. \ref{eqn:kappasum}, we obtain maps of lensing convergence.
Since each snapshot is obtained from the same initial conditions, to avoid artificial correlation,
we randomly shift and rotate the snapshots utilizing the periodical boundary condition.

We stack eight snapshots between $z=0$ and $z=1$ to make $\kappa$ maps.
These lensing maps correspond to source redshift $z\approx 1.02$ and comoving source distance $2400\mpch$.
The map size is $7.64^{\circ}\times 7.64^{\circ}$.
The maps all have $512^2$ uniform grids, corresponding to angular resolution $0.9^{'}$.
It is enough to resolve nonlinear regime.

This map making algorithm is valid under the Born approximation.
Higher-order lensing corrections \cite{Borncorrections} can be neglected in the present theoretical study.
Nonetheless, the process is still highly simplified in the sense that it ignores many complexities in real observations,
such as the random shape error in cosmic shear measurement, irregular masks and nonuniform survey depth.
We will briefly discuss the possible impact of these complexities later in the paper, but postpone quantitative study elsewhere.


\bfi{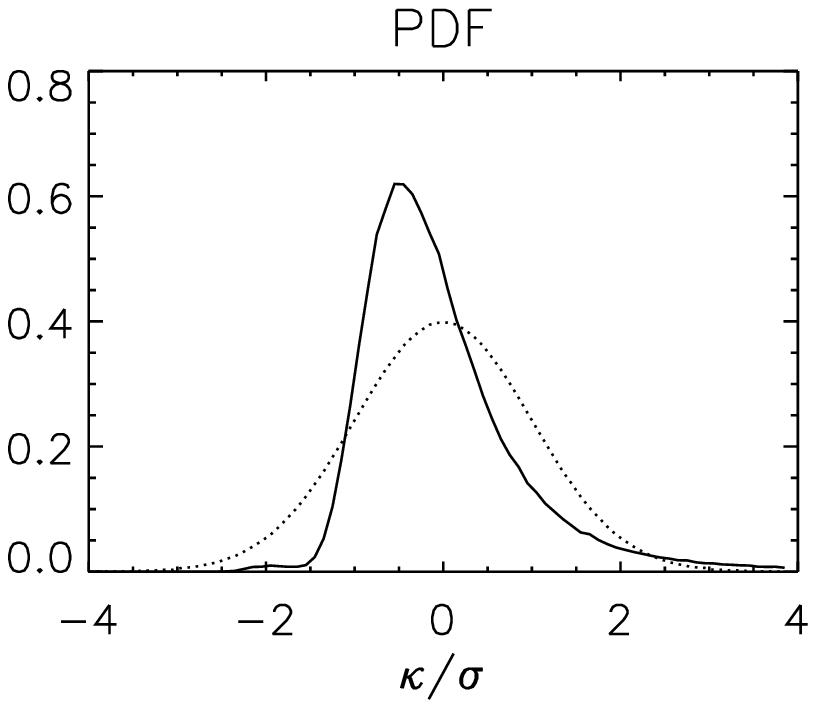}\\
\caption{The PDF of one $\kappa$ field (solid line) generated from our simulation.
The source redshift $z_S=1.02$ and the pixel size is $0.9^{'}$.
$\sigma$ is the rms dispersion in $\kappa$.
The dotted line is the Gaussian PDF.
Nonlinear evolution pushes the $\kappa$ PDF to be skewed toward negative value of $\kappa$,
while develops a tail of large positive $\kappa$.
\label{fig:lcpdf}}
\efi

\section{Gaussianization of the $\kappa$ field}
\label{sec:kappa}

It is well known that the lensing convergence field $\kappa$ is non-Gaussian at subdegree scale.
We show one such non-Gaussian PDF of a fine-scale $\kappa$ map we constructed in Fig.\ref{fig:lcpdf}.
Nonlinear evolution generates and amplifies non-Gaussianity in the matter distribution.
It drives $P(\kappa)$ to be skewed and shifts the peak to $\kappa<0$.

\bfi{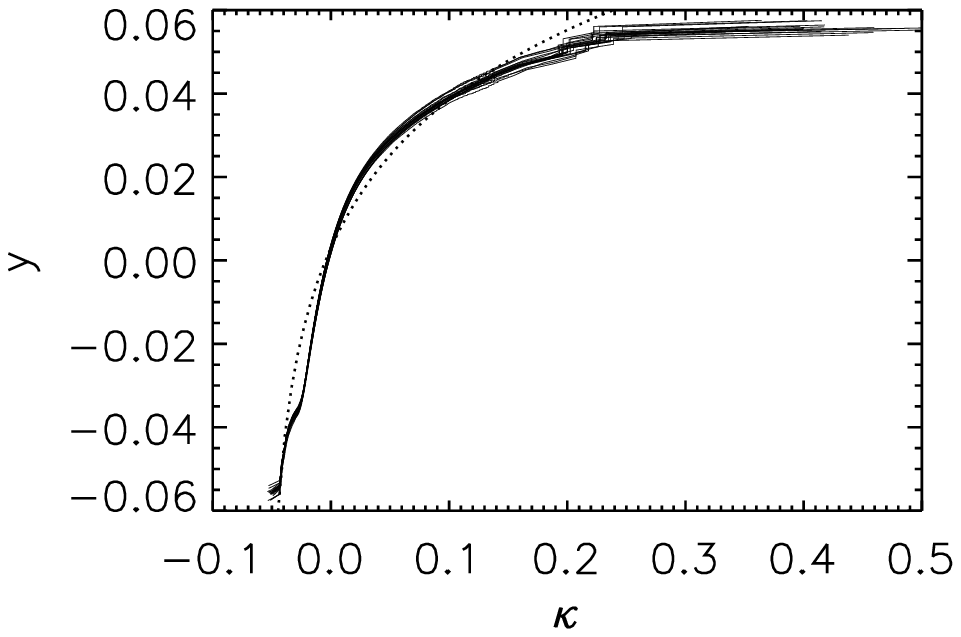}
\caption{The $\kappa$-$y$ relations for $20$ different line-of-sight realizations (solid line).
The Gaussian transformations show reasonably good agreement.
Irregularities at large $\kappa$ are largely caused by rare massive halos in the line-of-sight.
The dotted line is a fitting of generalized logarithmic form, $y=a\ln(1+\kappa/b)+c$, with $a$ and $b$, $c=y(\kappa=0)$ as fitting parameters.
We confirm that the logarithmic transformation is indeed a reasonably good approximation to the true Gaussianization transformation.
However, it is not surprising for the departures between them.
\label{fig:kappay}}
\efi

\subsection{The $\kappa$-$y$ relation}
\label{subsec:kappay}
The Gaussian transformations $\kappa \rightarrow y$ defined by Eq. \ref{eqn:kappay} are shown in Fig. \ref{fig:kappay},
for the 20  maps of $\kappa$ generated from our simulation.
The 20 curves for each realization (map) are in reasonable agreement with each other,
implying that the transformation is stable and insusceptible to field to field fluctuations \cite{footnote2}.
To fix the normalization, we require
\be
\label{eqn:kappaynormalization}
\left.\frac{{\dif}y}{{\dif}\kappa}\right|_{\kappa=0}=1\ .
\ee
by numerically estimating the slope at zero point of the standard Gaussian transformation function
(which make $y$ fields be standard Gaussian distributed with unit deviation), and scaling the slope to unity.
This condition makes the power spectrum at sufficiently linear scale more or less invariant under the transformation.
This condition is also satisfied by the usual logarithmic transformation  $\delta\rightarrow \ln (1+\delta)$ (e.g. in \cite{Neyrinck09,Neyrinck10}).
Nevertheless, our choice of the normalization is rather arbitrary.
Later in \S\ref{sec:conclusion} we will discuss whether we can use this normalization freedom
to not only Gaussianize but also {\it linearize} the matter density field.
Many more simulations are required to obtain the mean $\kappa$-$y$ relation, a key issue of further investigation.

How well can the widely adopted logarithmic transformation describe the above $\kappa$-$y$ relation?
Also plotted in Fig.\ref{fig:kappay} is a fitting to a generalized logarithmic form $y=a\ln(1+\kappa/b)+c$
with  $a$, $b$, $c$ as fitting parameters.
By definition, $c=y(\kappa=0)$. So in the fitting we simply fix $c$ to be $y(\kappa=0)$ averaged from 20 realizations.
The best fitting we found is $a=0.0145$, $b=0.0526$ and $c=0.00294$.
This fitting is reasonably good, confirming the usefulness of the logarithmic transformation.
However, Fig.\ref{fig:kappay} does show clear deviations from even a general logarithmic transformation defined above.
It is not surprising that the fields are not exactly lognormally distributed,
since the departures from lognormality have already been found by several works.

The Gaussian transformation suppresses $\kappa$ of high positive values relative to the one of lower values (Fig. \ref{fig:kappay}).
Since these high $\kappa$ are largely responsible for the non-Gaussianity,
we would expect the above transformation will suppress the overall non-Gaussianity to some extent.

\bfi{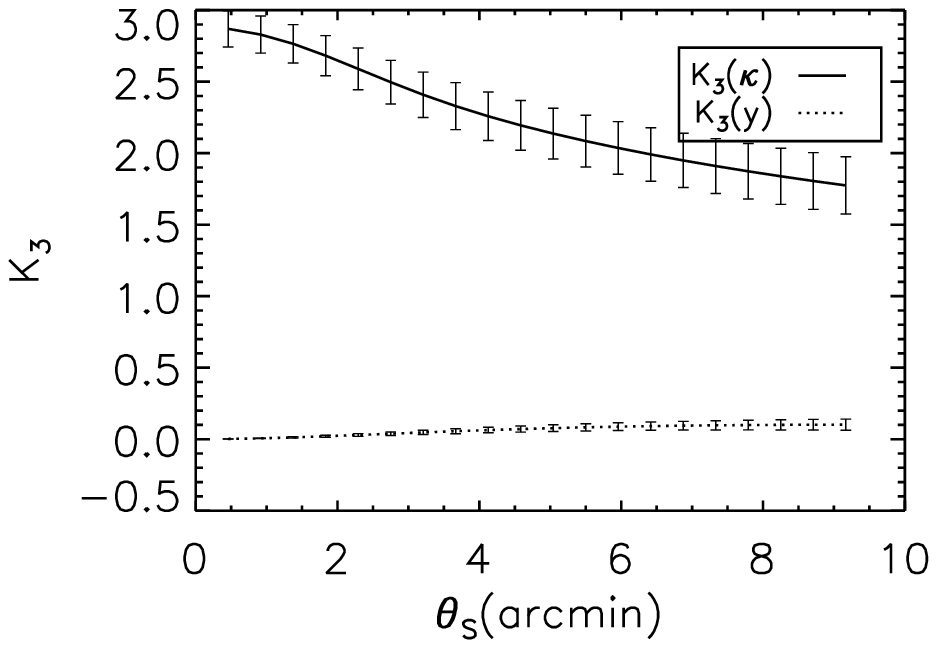}\\
\caption{The skewness $K_3$ of the $\kappa$ fields (solid line) and the $y$ fields (dotted line) smoothed with Gaussian filters of radius $\theta_S$.
The lines are averaged over 20 realizations and error bars drawn are the rms dispersions over the 20 realizations.
The cumulants $K_n$ ($n=3,\cdots$) are defined by Eq. \ref{eqn:cumulants}).
They have different normalizations from  that of the  cumulants $S_n$ widely adopted in the LSS literature.
Each $K_n$ is invariant under the scaling $y\rightarrow a y$ and hence avoid an ambiguity in quantifying the non-Gaussianity.
\label{fig:lcskewness}}
\efi

\bfi{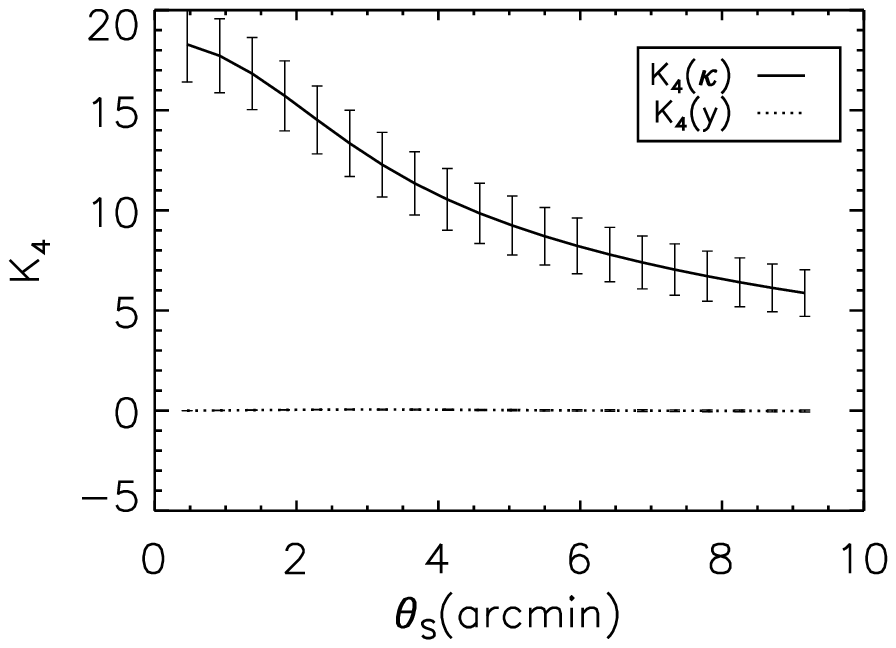}
\caption{Same as Fig. \ref{fig:lcskewness}, but for the kurtosis $K_4$.
The Gaussianization suppresses $K_4$ dramatically, so $K_4(y)$ is barely recognizable in this plot.
Refer to Fig. \ref{fig:lcs3rcmp} for a clearer view.
We caution the readers on the definition of $K_n$ (Eq. \ref{eqn:cumulants}). \label{fig:lckurtosis}}
\efi

\bfi{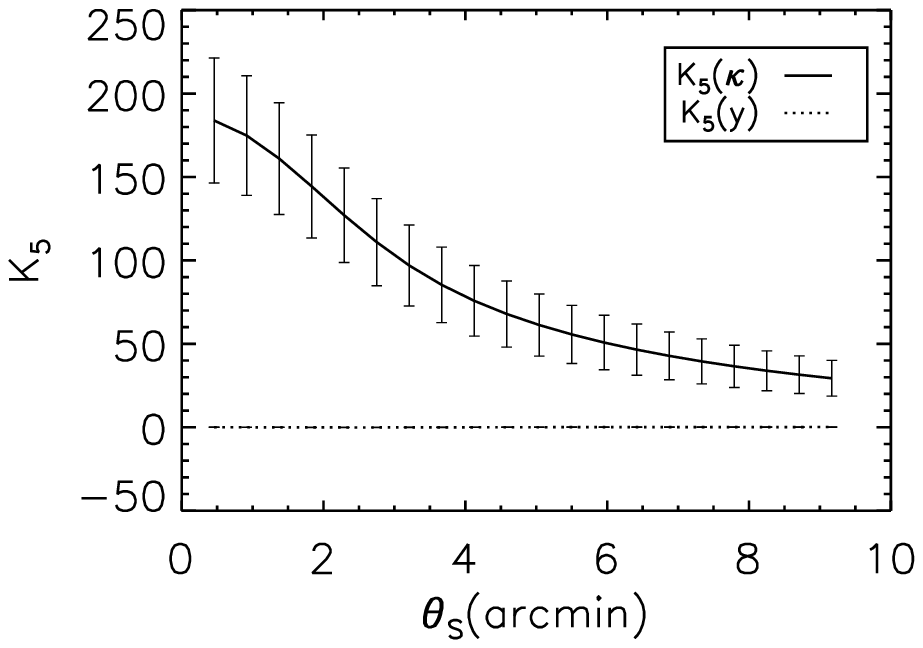}\\
\caption{Same as Fig. \ref{fig:lcskewness}, but for the 5th-order cumulants $K_5$.
$K_5$ is suppressed by the Gaussianization by two orders of magnitude to effectively zero.
Refer to Fig. \ref{fig:lcs3rcmp} for a clearer view.
We caution the readers on the definition of $K_n$ (Eq. \ref{eqn:cumulants}).
\label{fig:lc5thcumu}}
\efi

\bfi{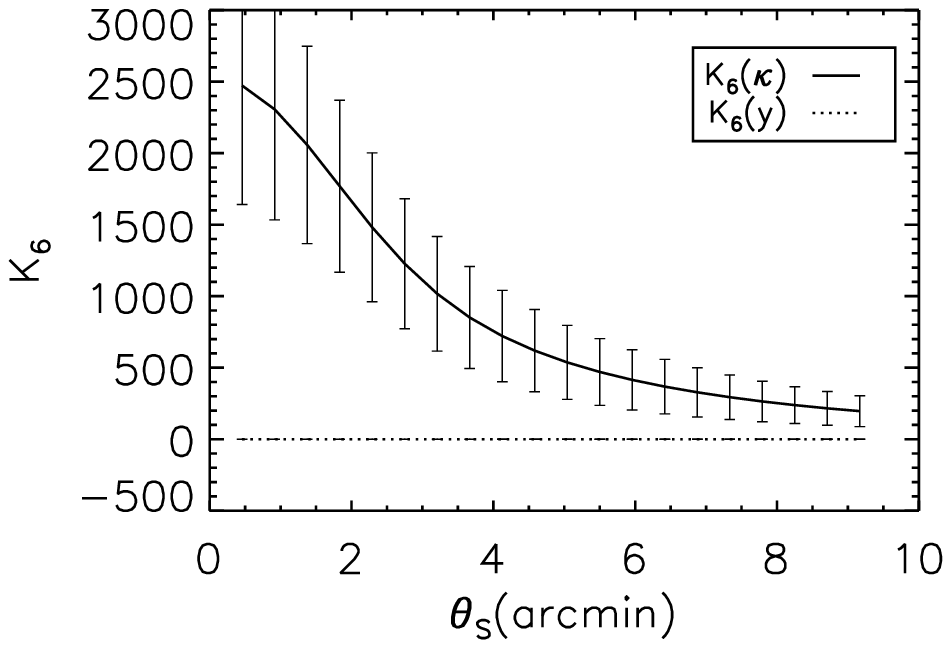}\\
\caption{Same as Fig. \ref{fig:lcskewness}, but for the 6th-order cumulants $K_6$.
We caution the readers on the definition of $K_n$ (Eq. \ref{eqn:cumulants}).
\label{fig:lc6thcumu}}
\efi

\bfi{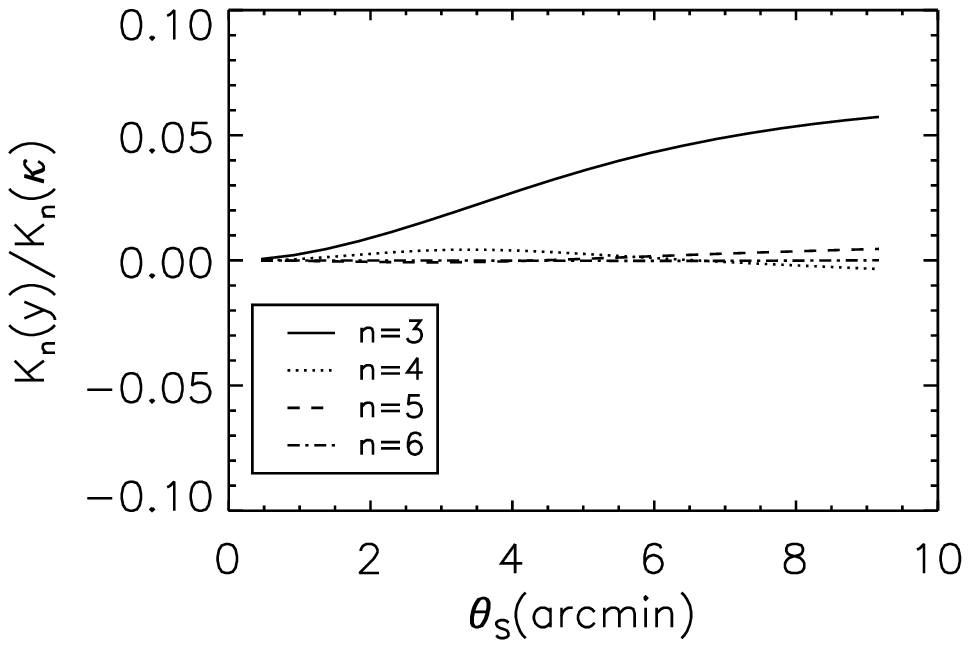}\\
\caption{The efficiency of the Gaussianization, in terms of the ratio of
high order cumulants before and after the Gaussian transformation.
The solid line is for the skewness, the dotted line for the kurtosis,
the dash line for the 5th-order cumulant, and the dotted dash line for the 6th-order
cumulant.  For all investigated scales, the cumulants are suppressed by at
least a factor of $\sim 20$. $K_n$, defined  through
Eq. \ref{eqn:cumulants}, is  invariant under the transformation
$y\rightarrow ay$, where $a$ is an arbitrary constant. This invariance makes
the ratio of the cumulants before and after the Gaussianization suitable for
quantifying the Gaussianization performance.
\label{fig:lcs3rcmp}}
\efi

\subsection{The cumulants}
\label{subsec:kappacumu}

The skewness $K_3$ of the smoothed $\kappa$ fields and $y$ fields, as a function of the smoothing radius $\theta_S$, are shown in Fig. \ref{fig:lcskewness}.
We adopt a Gaussian smoothing function, with $\theta_S\in [0.5^{'}, 9^{'}]$.
We somewhat arbitrarily choose the largest smoothing scale as $10$ pixel size.
Nevertheless, such scale is sufficient to take the pixel-pixel correlation into account and study the Gaussianization performance.
The Gaussianization procedure suppresses the skewness by at least an order of magnitude for all smoothing radius investigated.
The residual skewness at all $\theta_S$ is consistent with zero within three times the simulation error bars.
The performance at $\theta_S\rightarrow 0$ is trivial, other than verifying the accuracy of the solution to Eq. \ref{eqn:kappay}.
However, the performance at larger $\theta_S$ is indeed surprising.
The largest smoothing scale investigated is $\theta_S\sim 9^{'}$, corresponding to $10$ pixel size.
Smoothing over this radius includes complicated nonlocal correlation between pixels,
whose non-Gaussianity is not guaranteed to be removed or reduced by the local Gaussianization procedure.
Exactly what kind of feature in the LSS non-Gaussianity is responsible for the surprisingly good performance of the Gaussianization at large smoothing radius
is an interesting topic in itself and worth investigating.

Nevertheless, we find that the Gaussianization does not perfectly produce a Gaussian random field.
For example, $K_3$ at $\theta_S\sim 5^{'}$ shows weak deviation ($\sim 2\sigma$) from zero,
implying that the non-Gaussianity in the nonlocal pixel correlation is not completely removed by the Gaussianization.
The nonlocal pixel correlation is better described by the bispectrum,
in which we find more robust evidence on the residual non-Gaussianity (\S\ref{subsec:kappabisp}).

The Gaussianization also suppresses the kurtosis, the 5th and 6th order cumulants dramatically.
The residuals nearly vanish for all smoothing radius $\theta_S<10^{'}$ investigated (Fig. \ref{fig:lckurtosis},
\ref{fig:lc5thcumu} \&  \ref{fig:lc6thcumu}).
Fig. \ref{fig:lcs3rcmp} shows explicitly the suppression factors in cumulants up to the 6th order.
For all these quantities at all smoothing radius, they are suppressed by at least an order of magnitude.
All these results demonstrate the excellent performance of the Gaussianization procedure.
These results also provide strong and independent support of the Gaussian copula hypothesis in \cite{Scherrer10}.

\begin{figure*}
\epsfxsize=16cm
\epsffile{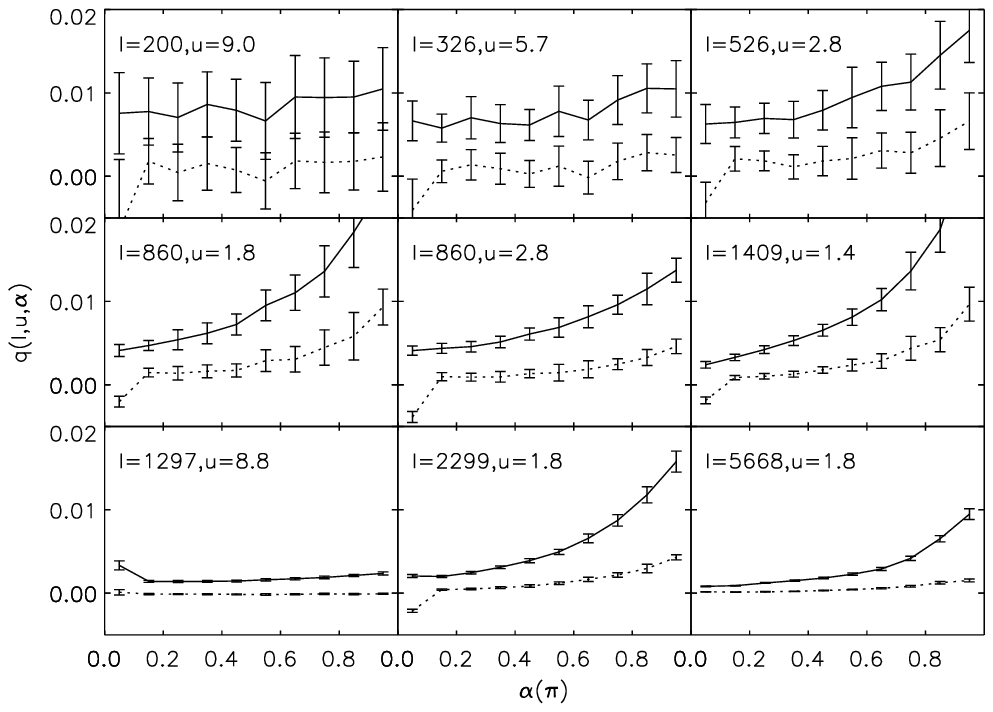} 
\caption{The redefined reduced bispectrum $q$s before (solid lines) and after (dotted lines) the Gaussianization.
The Gaussianization reduces the $q$s for almost all configurations and reduce many of them to be consistent with zero within $1\sigma$ error.
However, for some configurations the residual non-Gaussianity is still significant. }
\label{fig:lcbps}
\end{figure*}

\subsection{The bispectrum}
\label{subsec:kappabisp}

Another measure of non-Gaussianity is bispectrum $B(\vec{l}_1,\vec{l}_2, \vec{l}_3)$, which vanishes for Gaussian fields.
The bispectrum in LSS study is often expressed in a reduced bispectrum $Q(l_1,l_2,l_3)$ for convenience.
However, $Q$ defined in the conventional way is not invariant under the scaling $y\rightarrow a y$ whereas the non-Gaussianity should be.
Hence, following the same argument in defining the cumulants, we define a reduced bispectrum $q$ according to Eq. \ref{eqn:Q},
which is invariant under the transformation $y\rightarrow ay$.
We express $q$ in the coordinate $(l,u,\alpha)$, where $l\equiv l_1$, $u\equiv l_2/l_1$ and $ \alpha=\arccos(\vec{l}_1\cdot\vec{l}_2)/l_1 l_2$.

The reduced bispectrum for various configurations are presented in Fig.\ref{fig:lcbps},
where the solid lines are for the $\kappa$ fields and dotted are for the corresponding $y$ fields.
The curves are averaged over the 20 realizations and the error bars are the rms dispersions within the 20 realizations.
We show $9$ values of $l\in [200,6000)$, which covers most range of interest for weak lensing cosmology.
For most configurations of $(l, u,\alpha)$, $q$ is significantly suppressed by  the Gaussianization.
In many configurations the $q(y)$s are consistent with zero.
Since the bispectrum is a measure of nonlocal pixel correlation, these results show effectiveness of the Gaussianization.
This conclusion is consistent with the findings in previous section on the skewness and higher-order cumulants.

The $q$s (and the bispectra) of the $y$ field scatter about zero.
This behavior is in sharp contrast to the $q$s of the $\kappa$ field, which is  always positive, a characteristic behavior of the large scale structure.
The skewness is an average of the bispectrum over all configurations (although with different weighting).
Positive and negative values of the bispectrum of different configurations largely cancel in the average.
This explains the larger suppression in the skewness (Fig.\ref{fig:lcs3rcmp}) than in the bispectrum of individual configuration.

Since $q$ describes the non-Gaussianity of individual configuration, to some extent it is a more sensitive measure of the performance of the Gaussianization.
We find robust evidence for residual non-Gaussianity in some configurations of the bispectrum (e.g. some configurations with large $\alpha$ in Fig. \ref{fig:lcbps}).
This means that the non-Gaussian and nonlocal pixel correlation is not completely removed by the Gaussianization.
It also confirms the residual nonlocal pixel correlation as the source of residual non-Gaussianity.
A direct implication is that the weak lensing convergence copula is not exactly Gaussian, otherwise the bispectrum of the $y$ field would vanish.
Another implication is that perfect local Gaussianization transformation does not exist.


\bfi{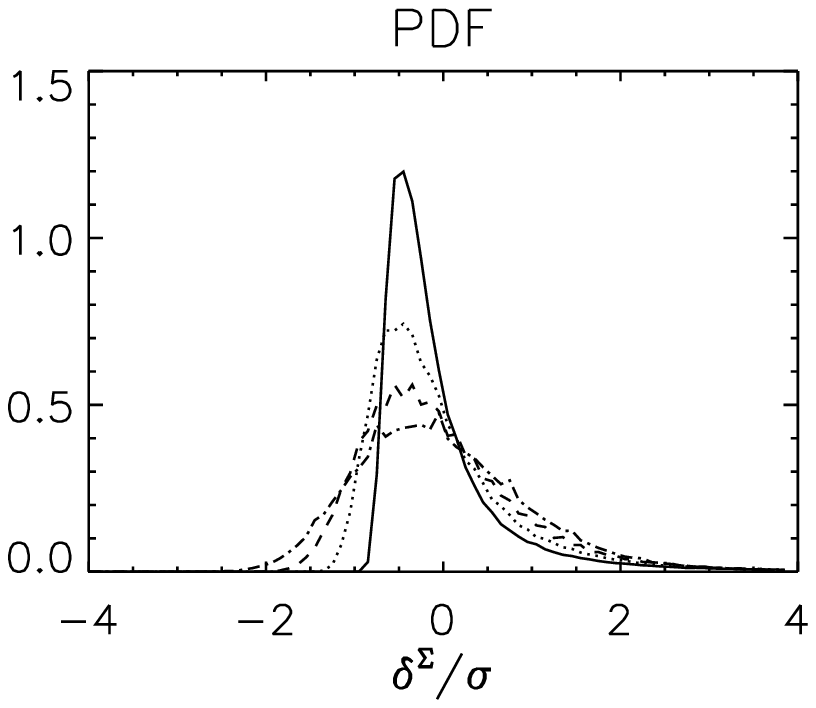}\\
\caption{The PDF of $\delta^\Sigma$ field at $z=0$, $1$, $2$, $3.5$. $\sigma$
  is the corresponding rms dispersion per pixel. The peak
  height increases with decreasing redshift.
\label{fig:pdf}}
\efi

\section{Gaussianization of the $\delta^\Sigma$ fields}
\label{sec:sigma}

Weak lensing tomography allows us to reconstruct the matter distribution as a function of redshift and hence infers the LSS evolution \cite{lensingtomography}.
For this reason, we investigate the Gaussianization of the $\delta^\Sigma$ field, the 2D matter density distribution projected over a narrow redshift bin.

We generate $\delta^\Sigma$ maps on a uniform $512^2$ grid at various redshifts.
The pixel size of the $\delta^\Sigma$ maps is then $L/512=0.6\mpch$.
The one-point PDFs of the fine-scale $\delta^{\Sigma}$ maps are shown in Fig. \ref{fig:pdf}.
Clearly, nonlinear evolution generates and amplifies non-Gaussianity in the matter distribution.
It drives $P(\delta^\Sigma)$ to be skewed and shifts the peak to $\delta^\Sigma<0$.
This is most obvious at $z\la 2$, the most relevant redshift range for weak lensing tomography.

\bfi{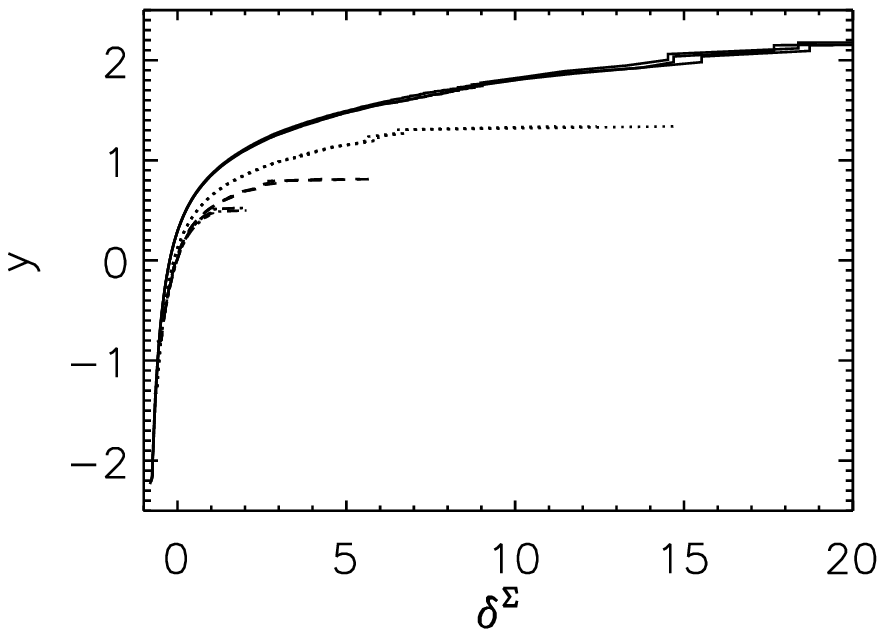}\\
(a)\\
\epsfxsize=8cm \epsffile{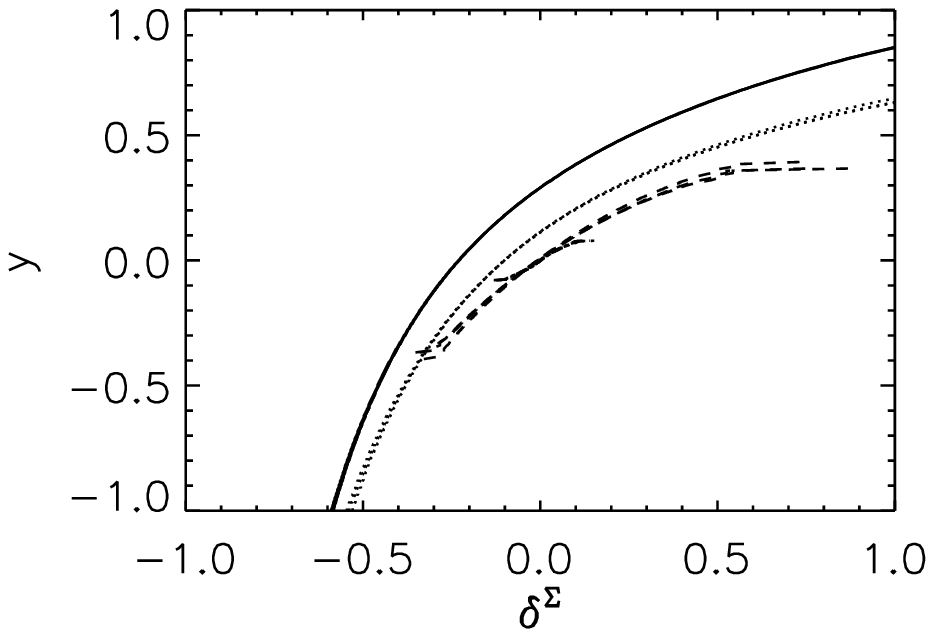}\\
(b)
\caption{Panel (a) shows the Gaussianization relation $\delta^\Sigma$-$y$.
From up to bottom, the lines correspond to $z=0,1,2,3.5$.
For each redshift, we plot the transformation functions obtained from 3 independent Cartesian directions,
which converge well and are barely distinguishable.
Panel (b) is zoom-in version near $\delta^\Sigma=0$.
Notice the shift of $y(\delta^\Sigma=0)$ and $\delta^\Sigma(y=0)$ with respect to the redshift.
\label{fig:xxrltn}}
\efi

\subsection{The $\delta^\Sigma$-$y$ relation}
\label{subsec:sigmay}

We will basically reapply the analysis done to the $\kappa$ field to the $\delta^\Sigma$ fields.
Again, we fix the normalization by setting
\be
\left.\frac{{\dif}y}{{\dif}\delta^\Sigma}\right|_{\delta^\Sigma=0}=1\ .
\ee
This condition is designed to keep the power spectrum at sufficiently linear scale invariant
under the transformation $\delta^\Sigma\rightarrow y$, a desirable property to study the structure evolution.

The Gaussian transformations $\delta^\Sigma$-$y$ at $z=0,1,2,3.5$ are shown in Fig. \ref{fig:xxrltn}.
To check the stability of the transformation, we plot the $\delta^\Sigma$-$y$ relations on the three independent projections of the same output.
Fig.\ref{fig:xxrltn} shows that the three curves for each output overlap with each other and  are virtually indistinguishable.
This convergence implies that the transformation is stable and insensitive to field to field fluctuations.

As expected, the Gaussian transformation suppresses $\delta^\Sigma$ of large amplitude relatively to $\delta^\Sigma$ of small amplitude.
This is also what we would expect for the logarithmic transformation $\delta^\Sigma\rightarrow \ln(1+\delta^\Sigma)$.
However, closer look shows that it differs from the logarithmic transformation.
(1) We have checked that $\ln(1+\delta^\Sigma)$-$y$ is not a straight line, with noticeable departures from the logarithmic transformation.
(2) The logarithmic transformation and its generalization $\delta^\Sigma\rightarrow \ln(1+\delta^\Sigma/b)$ contains a fixed point $(0,0)$.
Here $b$ is some constant.
But the actual Gaussian transformation does not have this fixed point.
The point $y(\delta^\Sigma=0)$ shifts toward larger positive value and the point $\delta^\Sigma(y=0)$ shifts toward larger negative value, for decreasing redshift.
This behavior is consistent with the picture of nonlinear evolution.
At sufficiently high redshifts, underdense ($\delta^\Sigma<0$) regions occupy half the cosmic volume.
Towards lower redshifts, the volume occupied by underdense regions (voids) increases.
This causes $y(\delta^\Sigma=0)$ to become increasingly positive.
Meanwhile, $\delta^\Sigma$ in the voids evolves to more negative value.
Since $\delta^\Sigma(y=0)$ is the threshold that voids with lower density occupy half the cosmic volume, it becomes increasingly negative with decreasing redshift.

\bfi{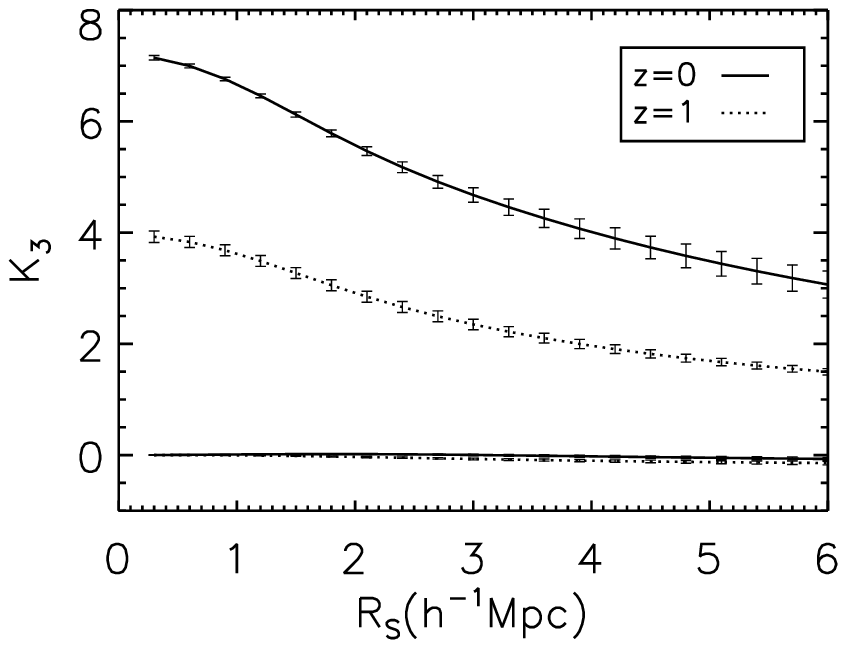}\\
(a)\\
\epsfxsize=8cm \epsffile{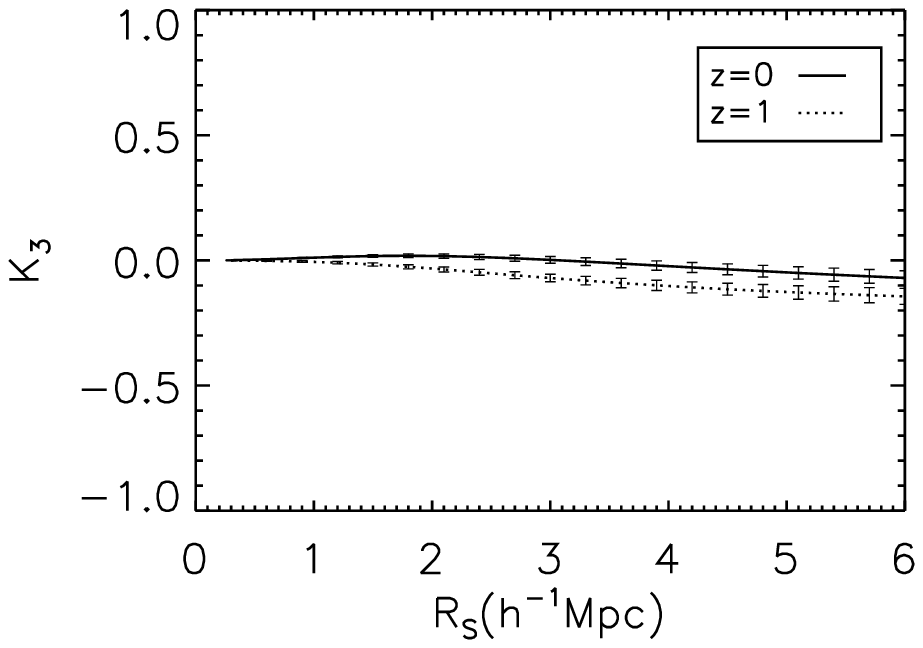}\\
(b)
\caption{Panel (a) shows the skewness of $\delta^\Sigma$ field and of $y$ field depending on smoothing radius
on averaging results of three directions of simulation box.
Solid lines for $z=0$ and dotted ones for $z=1$.
Errors are calculated through three different directions.
Panel (b) only shows the results of the $y$ field such that we can see the residual non-Gaussianity clearly.
\label{fig:skewness}}
\efi

\bfi{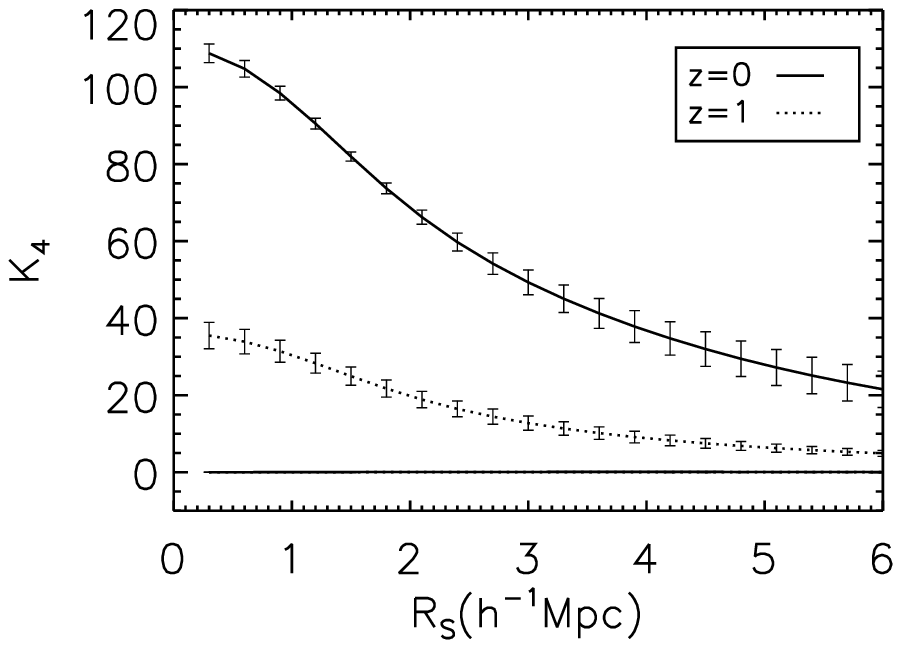}\\
(a)\\
\epsfxsize=8cm \epsffile{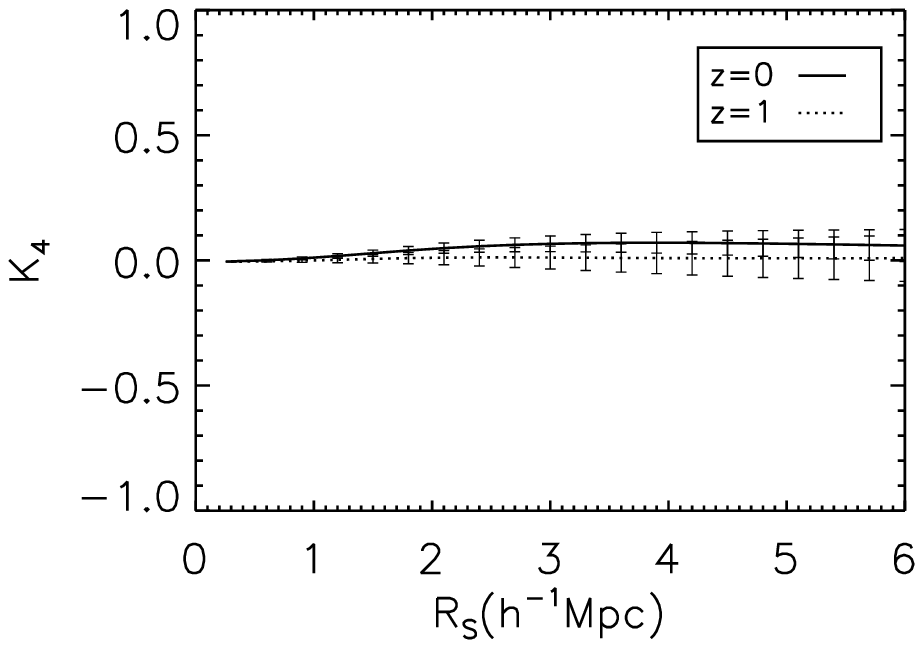}\\
(b)
\caption{Similar to Fig. \ref{fig:skewness}, but for $K_4$.
The Gaussianization significantly suppresses $K_4$, so the data points of $K_4(y)$ are not recognizable in these plots.
Refer to Fig.\ref{fig:mucmp} for a clearer view.
\label{fig:kurtosis}}
\efi

\bfi{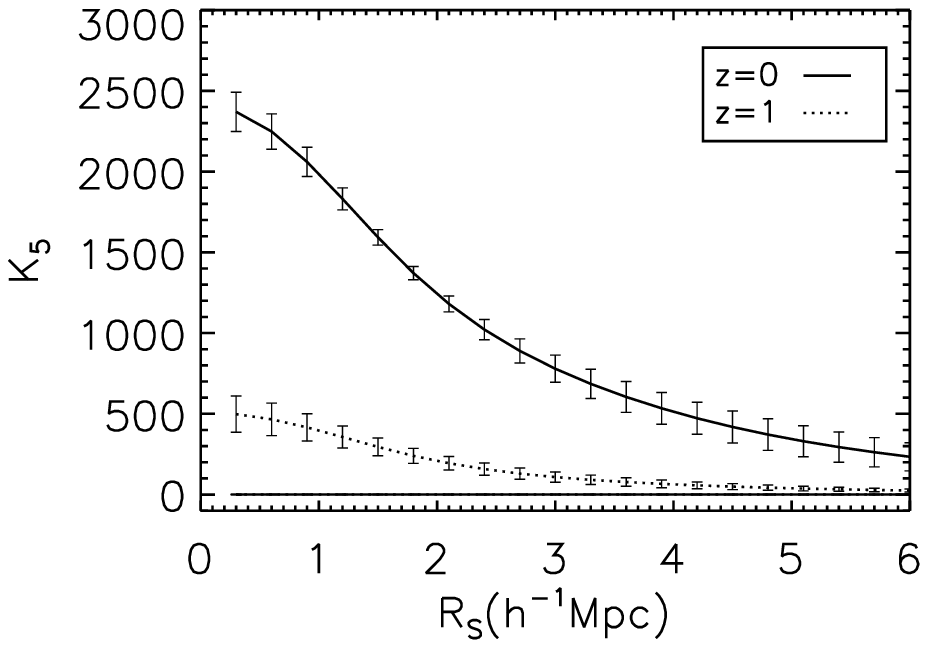}\\
(a)\\
\epsfxsize=8cm \epsffile{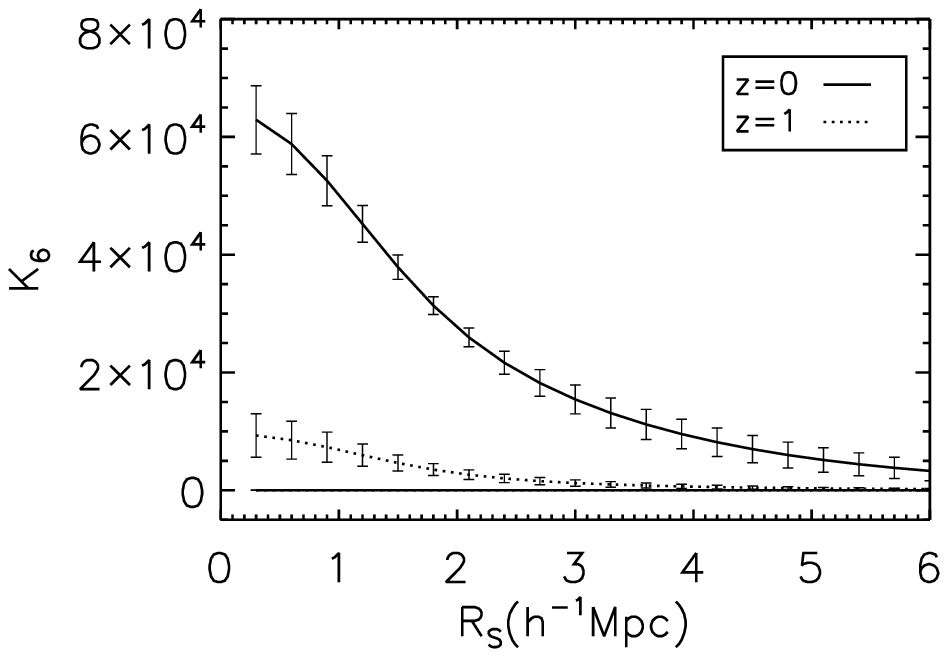}\\
(b)
\caption{Similar to Fig. \ref{fig:skewness}, but for $K_{5,6}$.
The Gaussianization significantly suppresses $K_{5,6}$, so the data points of $K_{5,6}(y)$ are not recognizable in these plots.
Refer to Fig.\ref{fig:mucmp} for a clearer view.
\label{fig:mun}}
\efi

\bfi{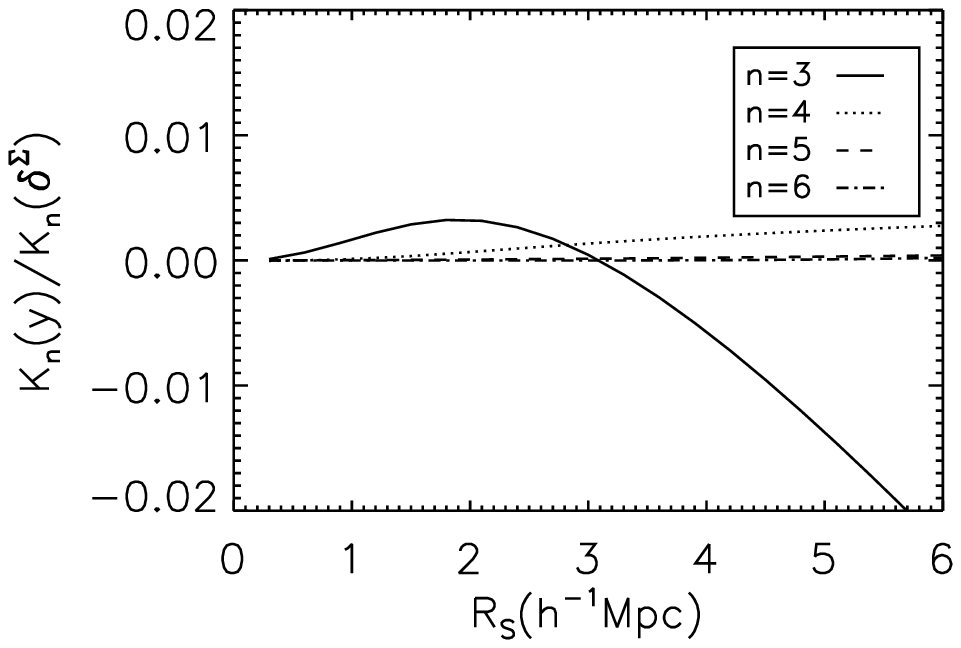}
\caption{The ratio of cumulants before and after the transformation as a function of smoothing scale at $z=0$.
Solid lines are for the skewness, dotted lines for the kurtosis, dash lines for $K_5$ and dotted dash lines for $K_6$.
The Gaussianization usually suppresses these non-Gaussian measures by two orders of magnitude or more.
\label{fig:mucmp}} \efi

\subsection{The cumulants of the smoothed $y(\delta^\Sigma)$ fields}
\label{subsec:sigmacumu}
We adopt the Gaussian window function with radius $R_S$ to smooth the $\delta^\Sigma$ field.
The adopted $R_S\in (0,6) \mpch$ covers most nonlinear scale of interest.
We show the results at two typical redshifts $z=0, 1$ in Fig. \ref{fig:skewness}, \ref{fig:kurtosis}, \ref{fig:mun} \& \ref{fig:mucmp}.
We find that the cumulants up to 6th order are all significantly suppressed with respect to the corresponding values before the Gaussianization.
We remind the readers again that these cumulants are defined in the usual language of statistics (Eq. \ref{eqn:cumulants}) and differ from those in LSS studies.

Fig.\ref{fig:skewness} shows the skewness $K_3$ before and after the Gaussianization.
The error bars are estimated from the three directions.
For all the redshifts and smoothing radius $R_S$ that we investigate, we find at least an order of magnitude or more suppression in $K_3$,
meaning that the non-Gaussianity is significantly suppressed.
Furthermore, $K_3$ after the Gaussianization is now $|K_3|\ll 1$, meaning that the field is indeed nearly Gaussian for the measure of $K_3$.

Despite this excellent performance of Gaussianization, it is not perfect.
The panel (b) of Fig.\ref{fig:skewness} shows small, but robust departure from $K_3=0$ at large smoothing radius, meaning a residual non-Gaussianity in the $y$ field.
Again, like the case of the convergence field, large smoothing radius means the inclusion of more and more non-Gaussian pixel-pixel correlations,
which are not necessarily removed by the local Gaussianization process.
And again, through the bispectrum measurement in \S \ref{subsec:sigmabisp}, we show more robustly the residual non-Gaussianity in the pixel-pixel correlation.
Nevertheless, the residual skewness is small.
So the Gaussianization procedure works effectively.

Fig. \ref{fig:kurtosis} shows the results of the kurtosis $K_4$ and Fig. \ref{fig:mun} shows the results of $K_5$ and $K_6$.
Fig. \ref{fig:mucmp} shows the ratio of $K_n$ ($n=3,4,5,6$) posterior to and prior to Gaussianization.
The Gaussianization suppresses $K_n$ ($n=3,4,5,6$) by  more than 1 order of magnitude (Fig. \ref{fig:mucmp}).
Furthermore, we have $|K_n|\ll 1$, meaning that the residual non-Gaussianity is indeed small and the $y$ fields are effectively Gaussian.

\begin{figure*}
\epsfxsize=16cm
\epsffile{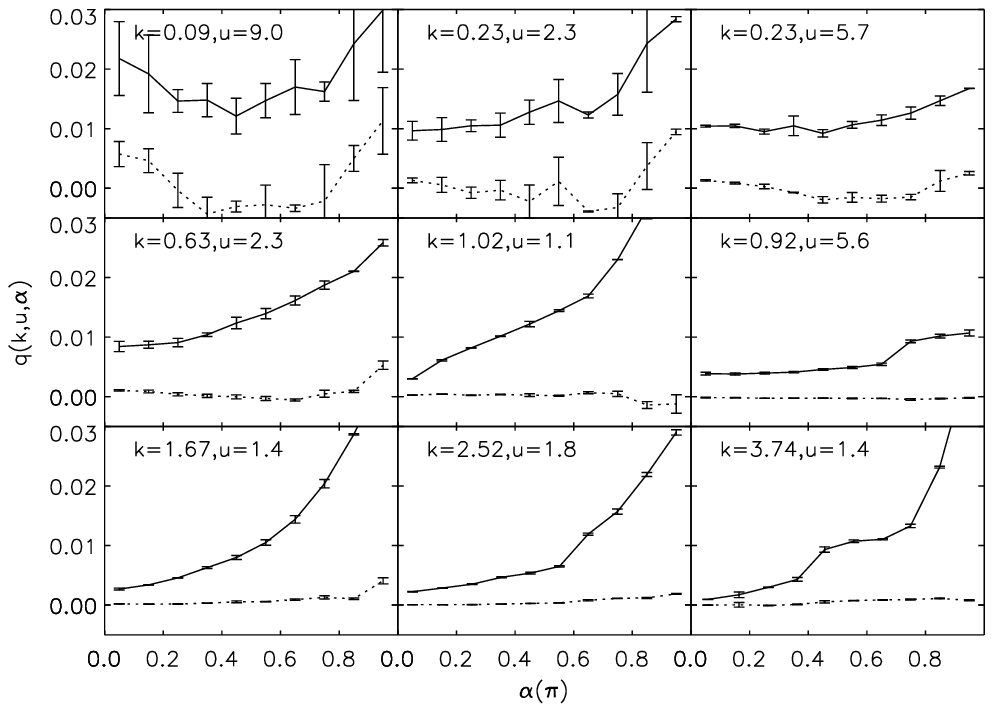} 
\caption{The redefined reduced  bispectrum $q$ of $\delta^\Sigma$ at $z=0$, before (solid lines) and after (dotted lines) the Gaussianization.
$k$ is in unit of $h/$Mpc.
Notice that $q$ differs from the widely adopted reduced bispectrum $Q$.
The Gaussianization suppresses the non-Gaussianity to effectively zero for almost all configurations.
However, we do find significant residual non-Gaussianity for some configurations, showing the imperfectness of the Gaussianization procedure.}
\label{fig:bps}
\end{figure*}

\subsection{The bispectrum of the $\delta^\Sigma$ field}
\label{subsec:sigmabisp}

The reduced bispectrum $q$ at $z=0$ is shown in Fig. \ref{fig:bps}.
Though $z=0$ is not the optimal redshift for the weak lensing tomography.
We choose this particular redshift because the non-Gaussianity is strongest in the $\delta^\Sigma$ field at $z=0$ but weakest in the $\kappa$ field.
By testing the Gaussianization performance for the two extreme cases, we can have better understanding on the generality of the Gaussianization procedure.

Again, the results are similar to the case of the convergence field.
Overall, the Gaussianization pushes $q$ toward zero for all configurations, a clear demonstration that Gaussianization works.
Furthermore, the bispectrum prior to Gaussianization is positive, whereas afterwards it is scattered about zero.
Since the skewness is a linear integral of the bispectrum, positive and negative bispectrum in the $y$ field partly cancel each other
and result in larger suppression in the skewness, as shown in Fig. \ref{fig:skewness} \& \ref{fig:mucmp}.

The suppression factor in $q$ depends on the strength of nonlinearity in the corresponding configuration.
The suppression factor is in general small, for small $k$ and large $\alpha$, which corresponds to weaker nonlinearity.
On the other hand, for larger $k$ and smaller $\alpha$ corresponding to stronger nonlinearity, $q$ can be suppressed by orders of magnitude.
This behavior is consistent with the suppression in skewness, which is smaller for larger smoothing radii.


\section{Discussions and conclusions}
\label{sec:conclusion}

We have demonstrated that it is indeed possible to Gaussianize the lensing convergence $\kappa$ field
and the underlying matter density field beyond one-point PDF,
through a simple local transformation which is defined only to Gaussianize the one-point PDF.
(1) The Gaussianization procedure utilizes a local monotonic transformation $\kappa\rightarrow y$
(or $\delta^\Sigma\rightarrow y$) to make one-point PDF of $\kappa$ (or $\delta^\Sigma$) Gaussian.
This is a well defined mathematical problem and we have numerically shown that the solution is stable.
(2) We quantified the performance of the Gaussianization beyond one-point PDF against various measures of the non-Gaussianity,
such as the skewness, kurtosis, 5th- and 6th-order cumulants of the smoothed fields, and the bispectrum.
We found that the Gaussianization works surprisingly well.
It suppresses the above non-Gaussianity measures by orders of magnitude and effectively reduce them to zero.
This implies that in many exercises we can treat the resulting $y$ field as Gaussian.
The Gaussianization procedure then compresses cosmological information in higher-order statistics of the original lensing fields into the power spectrum of $y$.
For this reason, analyzing weak lensing statistics can be significantly simplified.

Although we have tested the Gaussianization through many realizations of the $\kappa$ and $\delta^\Sigma$ fields,
all of them are generated from a single simulation of finite box size and number of particles.
So our findings are inevitably affected by numerical issues such as cosmic variance, mass and force resolution, shot noise and aliasing effect \cite{Jing05}.
We do not expect that any of these factors will alter the major conclusion on the effectiveness of the Gaussian transformation.
Nevertheless, these numerical issues could in particular have larger impact on the residual non-Gaussianity,
which itself is weak and relatively hard to measure accurately.
We can imagine that these numerical issues  introduce error in $P(\kappa)$.
Such error then nonlinearly propagates into the $\kappa\rightarrow y$ relation
and may then cause non-negligible under- or overestimation of the residual non-Gaussianity in the $y$ field \cite{footnote3}.
However, it is very unlikely that the detected residual non-Gaussianity can be canceled exactly.
In this sense, the detection of the residual non-Gaussianity is robust, although we need many more simulations to measure the amplitude to high precision.

There are many key issues to be explored in future study. An incomplete list
includes the following.
\bi
\item The applicability to real data.
  So far we only consider a highly idealized case, which is suitable for the purpose of theoretical modeling of weak lensing.
  The same Gaussianization in principle should be also applicable to real data.
  But application is complicated by various measurement errors in real data.
  How well does the Gaussianization work with the presence of measurement errors is a crucial issue for further investigation.
  One of such complexities that we will consider is the random shape error.
  As long as the pixel size is sufficiently large, the central limit theorem drives its distribution to be Gaussian.
  Since the Gaussianization that we have proposed is nonlinear, it will render this Gaussian noise into a non-Gaussian one.
  Even worse, the same nonlinear transformation could mix the lensing signal and measurement noise.
\item The redshift evolution.
  We have shown the effectiveness of the Gaussianization to reduce the non-Gaussianity in the redshift resolved $\delta^\Sigma$ field.
  But it does not necessarily make the resulting $y$ field evolve linearly.
  Further investigations are needed to see if a choice of suitable scaling $y\rightarrow ay$ will result $y$ in a linearly evolving field.
\item Understanding the $\kappa\rightarrow y$ relation.
  We need many more simulation realizations to robustly calculate and understand this relation and its dependence on cosmology, pixel size, redshift and other factors.
  It will be interesting to explore the cosmological information encoded in this relation.
\item The residual non-Gaussianity.
  The proposed Gaussianization works surprisingly well, but it does not perfectly produce a Gaussian random field.
  The residual non-Gaussianity we have detected is weak and is unlikely to carry a significant amount of cosmological information.
  Nevertheless, it may be still worth investigating the cosmological information carried by the residual non-Gaussianities.
\ei

2D projected matter distribution $\delta^\Sigma$ is the projection of 3D matter distribution along one direction,
and weak lensing convergence field $\kappa$ is a summation of $\delta^\Sigma_i$ with different weight function.
Since copula describe the structure of the random field and how the one-point PDF join together to get n-point joint PDF,
we hope that projection and summation leaving copula Gaussian could be proofed.
Since galaxy distribution is a biased tracer of matter distribution, it seems to keep rank order.
Hence we hope that galaxy distribution also has Gaussian copula.
We expect that Copula will be a useful tool to approach high-order statistics.
There have already been some works in cosmology using copula.
Sato et al. 2010 (a)\&(b) \cite{Sato10a}\cite{Sato10b} use Gaussian Copula and $\chi^2$ one-point PDF to describe the power spectrum estimation.
They found that the likelihood derived in this way is more similar to the likelihood directly derived from simulation,
than Gaussian likelihood hypothesis usually made in cosmological parameter estimation pipeline.
Hence the parameter constrain is more accurate and the degeneracy is expected to be broken.

\section*{Acknowledgment}
We thank Scott Dodelson, Ue-Li Pen, Jun Zhang, Yi Zheng for useful discussions.
We also thank Pascal Elahi for useful suggestions and proofreading.
P.J.Z. thanks  the support of the one-hundred talents program of the Chinese academy of science,
the national science foundation of China (grant No. 10821302, 10973027 \& 11025316),
the CAS/SAFEA International Partnership Program for  Creative Research Teams and the 973 program grant No. 2007CB815401.
W.P.L. acknowledges the supports from Chinese National 973 project (No. 2007CB815401), Chinese National 863 project (No. 2006AA01A125),
NSFC project (10873027, 10821302, 10533030), and the Knowledge Innovation Program of the Chinese Academy of Sciences (grant KJCX2-YW-T05).
W.G.C. acknowledges a fellowship from the European Commission's Framework Programme 7,
through the Marie Curie Initial Training Network CosmoComp (PITN-GA-2009-238356).


\begin{thebibliography}{}
\bibitem{footnote1}
As pointed out in \cite{Scherrer10}, a Gaussian field has both a Gaussian copula and a Gaussian one-point PDF.
However, it is possible for a non-Gaussian field to have a Gaussian copula (e.g., any local monotonic transformation on a Gaussian field).
The Gaussian Copula Hypothesis proposed by \cite{Scherrer10} and the lognormality of one-point PDF of density field imply that
under logarithmic transformation, the new density field has a Gaussian one-point PDF
and a Gaussian Copula (which is invariant under the transformation).
That is to say, the new field is Gaussian, and higher-order correlations vanish.
Since the deviation from lognormality has already been found,
the logarithmic transformation can not fully Gaussianize the density field even though the copula is Gaussian.
\bibitem{footnote2}
We do notice field-to-field scatters in the $\kappa$-$y$ relation,
especially at both ends of large positive and negative values of $y$.
However, we find that this is largely due to numerical effect in the normalization condition Eq.\ref{eqn:kappaynormalization}.
It causes the normalization in $y$, quantified by the r.m.s dispersion $\sigma_y$, to vary from one map to another.
We have checked that, if we choose the same normalization in $y$, namely the same $\sigma_y$,
the 20 curves share virtually the same maximum and minimum, and they do overlap and become nearly indistinguishable.
The normalization factor only scales the amplitude of the power spectrum,
and does not change the non-Gaussianity measures result we used.
The $\kappa$-$y$ relation shows step functionlike irregularities at large $\kappa $ in Fig. \ref{fig:kappay}.
It is likely caused by rare massive halos. Also likely for this reason,
the positions of the irregularities vary from map to map.
\end{thebibliography}
\end{document}